\documentclass[12pt]{article}
\usepackage{graphicx}
\usepackage{authblk}
\usepackage[margin=1in]{geometry}
\usepackage{amsmath,amssymb}
\usepackage{siunitx}
\usepackage{graphicx}
\usepackage{booktabs}
\usepackage{enumitem}
\usepackage{hyperref}
\usepackage[capitalise]{cleveref}
\usepackage{subcaption}
\usepackage{xcolor}
\usepackage{cite}
\usepackage{hyperref}
\usepackage{float}
\usepackage{comment}
\usepackage{multirow}

\title{\textbf{Toward Generalizable Surrogate Models for Molecular Dynamics via Graph Neural Networks}}

\author[1]{Judah Immanuel}
\author[2]{Avik Mahata\thanks{Corresponding author: \textit{mahataa@merrimack.edu}}}
\author[3]{Aniruddha Maiti}

\affil[1]{Department of Computer Science, Merrimack College, North Andover, MA, USA}

\affil[2]{Department of Mechanical and Electrical Engineering, Merrimack College, North Andover, MA, USA}

\affil[3]{Department of Mathematics, Engineering, and Computer Science,\\
West Virginia State University, Institute, WV, USA}

\date{}

\begin{document}
\maketitle

\begin{abstract}
We present a graph neural network (GNN) based surrogate framework for molecular dynamics simulations that directly predicts atomic displacements and learns the underlying evolution operator of an atomistic system. Unlike conventional molecular dynamics, which relies on repeated force evaluations and numerical time integration, the proposed surrogate model propagates atomic configurations forward in time without explicit force computation. The approach represents atomic environments as graphs and combines message-passing layers with attention mechanisms to capture local coordination and many-body interactions in metallic systems. Trained on classical molecular dynamics trajectories of bulk aluminum, the surrogate achieves sub angstrom level accuracy within the training horizon and exhibits stable behavior during short- to mid-horizon temporal extrapolation. Structural and dynamical fidelity are validated through agreement with reference radial distribution functions and mean squared displacement trends, demonstrating that the model preserves key physical signatures beyond pointwise coordinate accuracy. These results establish GNN-based surrogate integrators as a promising and computationally efficient complement to traditional molecular dynamics for accelerated atomistic simulations within a validated regime.
\end{abstract}

\section{Introduction}

Molecular dynamics (MD) simulation is a deterministic, equation-of-motion--based framework rooted in classical statistical mechanics. By explicitly integrating Newton’s laws for interacting atoms, MD establishes a direct connection between microscopic dynamics in phase space and emergent macroscopic material behavior. Owing to this first-principles dynamical foundation, MD has been widely adopted across physics, chemistry, and engineering to study thermodynamic, structural, and kinetic phenomena at the atomic scale \cite{allen_tildesley, frenkel_smit, rapaport, badar2022molecular}. By numerically solving Newton’s equations of motion for large assemblies of atoms, MD simulations enable prediction of thermodynamic properties, elucidation of atomistic mechanisms underlying material failure, and direct observation of complex processes such as diffusion, phase transformations, crystal growth, and interfacial evolution \cite{plimpton, anwar2011uncovering}. In metallic systems and alloys in particular, MD has proven highly effective for investigating solidification pathways, defect nucleation, plastic deformation, and fracture, complementing both experiments and continuum-scale models \cite{mishin1999interatomic, bitzek_deformation, mahata2022modified, mahata2018understanding, mahata2024bridging}. Despite its broad applicability and conceptual simplicity, the practical utility of classical MD is fundamentally constrained by a tradeoff between accuracy and computational cost. Two primary bottlenecks dominate this limitation. The first is the fidelity of the interatomic potential, or force field. While \textit{ab initio} MD methods based on density functional theory (DFT) offer higher accuracy and explicit treatment of electronic structure, their computational expense restricts simulations to systems containing at most a few hundred atoms and timescales on the order of picoseconds \cite{marx_hutter, kresse_furthmuller}. The second, and more persistent, limitation arises from the intrinsic timescale separation in atomic motion. To maintain numerical stability, MD simulations require femtosecond-scale integration time steps (typically $\sim$1--5~fs) to resolve the fastest vibrational modes. As a result, accessing experimentally relevant timescales---ranging from nanoseconds to milliseconds, demands billions of integration steps, even when employing simple empirical potentials, leading to prohibitive computational costs \cite{kadau2006molecular, md_timescale_limit, nguyen2021billion}. Overcoming these accuracy and timescale barriers remains a central challenge in extending MD to mesoscale and realistic processing conditions.

In condensed matter systems, atomic interactions are predominantly local, governed by short-range forces arising from electronic screening and bonding. In MD simulations, this locality is explicitly enforced through neighbor lists and cutoff radii ($r_c$), beyond which interactions are neglected. While this representation is natural from a physical standpoint, translating irregular, continuous atomic configurations into effective machine learning models presents a significant topological challenge. Conventional machine learning architectures struggle to respect the fundamental symmetries and variable connectivity inherent to atomistic systems. Convolutional neural networks (CNNs), including three-dimensional variants, require mapping atomic positions onto fixed Euclidean grids via voxelization. This discretization leads to sparse, high-dimensional inputs, increased computational cost, and inefficient handling of rotational symmetries \cite{ryczko2018convolutional, kajita2017universal, peivaste2024rapid}. Recurrent neural networks (RNNs) \cite{schmidt2019recurrent}, on the other hand, impose an artificial sequential ordering on atomic environments, violating permutation invariance with respect to atom indexing. More traditional approaches, such as kernel methods or feed-forward neural networks, rely on fixed-length, handcrafted descriptors (e.g., atom-centered symmetry functions or SOAP descriptors), which decouple representation learning from model optimization and limit scalability to complex, variable-sized systems \cite{behler2007generalized, bartok2013soap}. Graph neural networks (GNNs) provide a natural and physically consistent alternative. In a GNN representation, atoms are treated as nodes and interatomic interactions within a cutoff radius are encoded as edges, yielding a flexible, non-Euclidean graph topology that directly mirrors the structure used in MD simulations. Through message-passing operations, each atom updates its latent state based only on information from neighboring atoms, enforcing locality, permutation invariance, and extensibility by construction \cite{gilmer2017mpnn}. This framework enables GNNs to model variable coordination environments without imposing artificial constraints on system size or topology. Over the past decade, machine learning has emerged as a powerful tool for accelerating materials modeling and discovery \cite{butler2018mlreview}. The most mature integration of machine learning with MD has focused on addressing the accuracy bottleneck through machine-learned interatomic potentials (MLIPs). Beginning with the seminal work of Behler and Parrinello \cite{behler2007generalized}, MLIPs have evolved rapidly, with GNN-based architectures proving particularly effective. Models such as SchNet \cite{schuett2017schnet}, the $E(3)$-equivariant NequIP \cite{batzner2022nequip}, and MACE \cite{batatia2022mace} can now reproduce DFT energies and forces with near-chemical accuracy while achieving orders-of-magnitude speedups relative to \textit{ab initio} MD. Despite these advances, MLIPs alone do not overcome the fundamental timescale limitation imposed by femtosecond integration steps. Even with near-instantaneous force evaluations, long-time simulations remain prohibitively expensive. This limitation has motivated a more ambitious direction: using machine learning to replace or augment the numerical time integrator itself. Rather than predicting forces at each step, GNN-based dynamical models can be trained to directly propagate atomic states forward in time, effectively learning a surrogate for the MD evolution operator. Early demonstrations on molecular and coarse-grained systems have shown that such models can achieve stable long-horizon predictions while preserving key physical invariants \cite{sanchez2020learning, zheng2021mdnet}. These approaches suggest a promising pathway toward bypassing the traditional time-stepping bottleneck and extending atomistic simulations to previously inaccessible spatiotemporal regimes. GNNs are uniquely suited for this task. An atomic configuration is a natural graph, and a GNN can learn the complex many-body interactions and spatial relationships that govern the system's dynamics. By operating directly on this graph representation, a GNN can learn a high-dimensional function that maps the system's state at time $t$ to its state at time $t + \Delta t$.

GNNs have increasingly been integrated with MD simulations for a variety of complementary purposes. Most prominently, they have been used to construct MLIPs that reproduce \textit{ab initio} energies and forces at a fraction of the computational cost. Starting from early neural network potentials \cite{behler2007generalized, li2022graph, park2021accurate, hofstetter2022graph}, modern GNN-based models such as SchNet \cite{schuett2017schnet}, DeePMD \cite{wang2018deepmd}, NequIP \cite{batzner2022nequip}, and MACE \cite{batatia2022mace} have demonstrated near-DFT accuracy across a wide range of materials systems. Beyond force-field construction, GNNs have also been applied to predict local atomic properties, bonding environments, reaction pathways, and charge distributions, enabling mechanistic insight into chemical bonding and structural evolution \cite{gilmer2017mpnn, smith2017ani, unke2019physnet}. More recently, GNNs have been explored as data-driven dynamical models capable of learning short-time atomic evolution directly from trajectory data. These approaches aim to approximate the MD time-propagation operator itself, rather than merely supplying forces to a classical integrator. Early demonstrations using graph-based simulators showed that GNNs can learn physically meaningful dynamics in molecular and coarse-grained systems while respecting permutation invariance and locality \cite{sanchez2020learning}. Subsequent work, such as MDNet \cite{zheng2021learning}, further established the feasibility of learning atomic displacements or velocities over finite time horizons. Despite these promising developments, existing approaches remain largely limited to short rollout lengths, simplified chemistries, or narrowly defined thermodynamic conditions. Consequently, several critical gaps persist in the current literature. First, the vast majority of ML--MD integrations focus on accelerating force evaluations while retaining classical numerical integration schemes, leaving the fundamental femtosecond time-step bottleneck unaddressed. Second, stable long-horizon propagation remains challenging due to error accumulation, drift in conserved quantities, and degradation of thermodynamic fidelity. Third, generalization across temperatures, strain states, defect populations, and chemical compositions is still poorly understood, with many models requiring extensive system-specific retraining. Finally, uncertainty quantification and physically grounded constraints---such as conservation laws, equivariance, and stability guarantees---are only beginning to be incorporated into GNN-based dynamical models. These limitations motivate the development of GNN-based surrogate models that replace the conventional MD integration loop altogether. By directly learning a high-dimensional mapping from the atomic state at time $t$ to the state at $t + \Delta t$, such models have the potential to bypass explicit force computation and numerical integration while preserving the essential physics of many-body interactions. When combined with physically informed architectures (e.g., equivariance, locality, and symmetry preservation), uncertainty-aware training, and adaptive temporal resolution, GNN surrogates offer a promising pathway toward extending atomistic simulations to experimentally relevant spatiotemporal scales that are inaccessible to conventional MD.

In this work, we develop a high-fidelity GNN surrogate model that directly predicts atomic displacements in metallic systems. Unlike conventional MD, which computes trajectories through repeated force evaluations followed by numerical time integration, the proposed approach learns the atomic evolution operator itself. Given the atomic configuration at time $t$, including positions, velocities, and local structural information, the model predicts the displacement $\Delta \mathbf{r}$ over a prescribed time interval $\Delta t$, thereby bypassing both explicit force calculations and classical integration schemes. The model employs a hybrid architecture that combines deep message-passing layers with transformer-style self-attention blocks, enabling it to capture local many-body interactions while also learning longer-range correlations within the atomic environment. This design yields a rich, physics-informed latent representation that respects permutation invariance and locality while remaining flexible enough to model complex metallic bonding environments. The network is trained on a comprehensive dataset generated from classical MD simulations of pure aluminum, comprising atomic positions, velocities, and structural descriptors sampled across a wide range of configurations. We demonstrate that the proposed GNN surrogate can predict atomic trajectories with sub-\AA{}ngstr{\"o}m accuracy over single-step and short-horizon rollouts, effectively functioning as a high-fidelity integrator for metallic systems. By learning atomic displacements directly, the model eliminates the dominant computational bottlenecks associated with force-field evaluation and femtosecond-scale time stepping. We position this work as a proof of concept for a new class of accelerated MD engines, in which GNN-based surrogates replace the traditional integration loop while retaining atomistic resolution and physical fidelity.


\section{Computational Methodology}
\subsection{Molecular Dynamics Simulation for Initial Data Generation}

To generate the initial dataset for the GNN, a series of classical MD simulations were performed on bulk aluminum using the \textsc{LAMMPS} simulation package \cite{thompson2022lammps}. These simulations provided equilibrated atomic configurations, velocities, and thermodynamic data across a broad temperature range, forming the foundation for the data-driven prediction of atomic displacements.

\subsubsection{Interatomic Potential and Theoretical Framework.}
The atomic interactions were modeled using the embedded atom method (EAM) potential \cite{daw1993embedded}, which accurately describes the many-body nature of metallic bonding. Within the EAM formalism, the total potential energy of a system is expressed as
\begin{equation}
E_{\text{total}} = \sum_i F_i(\rho_i) + \frac{1}{2} \sum_{i \ne j} \phi_{ij}(r_{ij}),
\end{equation}
where $F_i(\rho_i)$ is the embedding energy required to place atom $i$ into the background electron density $\rho_i$, and $\phi_{ij}(r_{ij})$ represents a pairwise interaction between atoms $i$ and $j$ separated by a distance $r_{ij}$. 
This approach captures the cohesive and metallic characteristics of aluminum while retaining computational efficiency suitable for large-scale simulations. 
The EAM potential effectively bridges quantum-mechanical effects with classical mechanics by implicitly accounting for local electron density redistribution, thereby providing a realistic description of metallic cohesion.

\subsubsection{Simulation Setup}
\begin{figure}[H]
    \centering
    \includegraphics[width=0.95\textwidth]{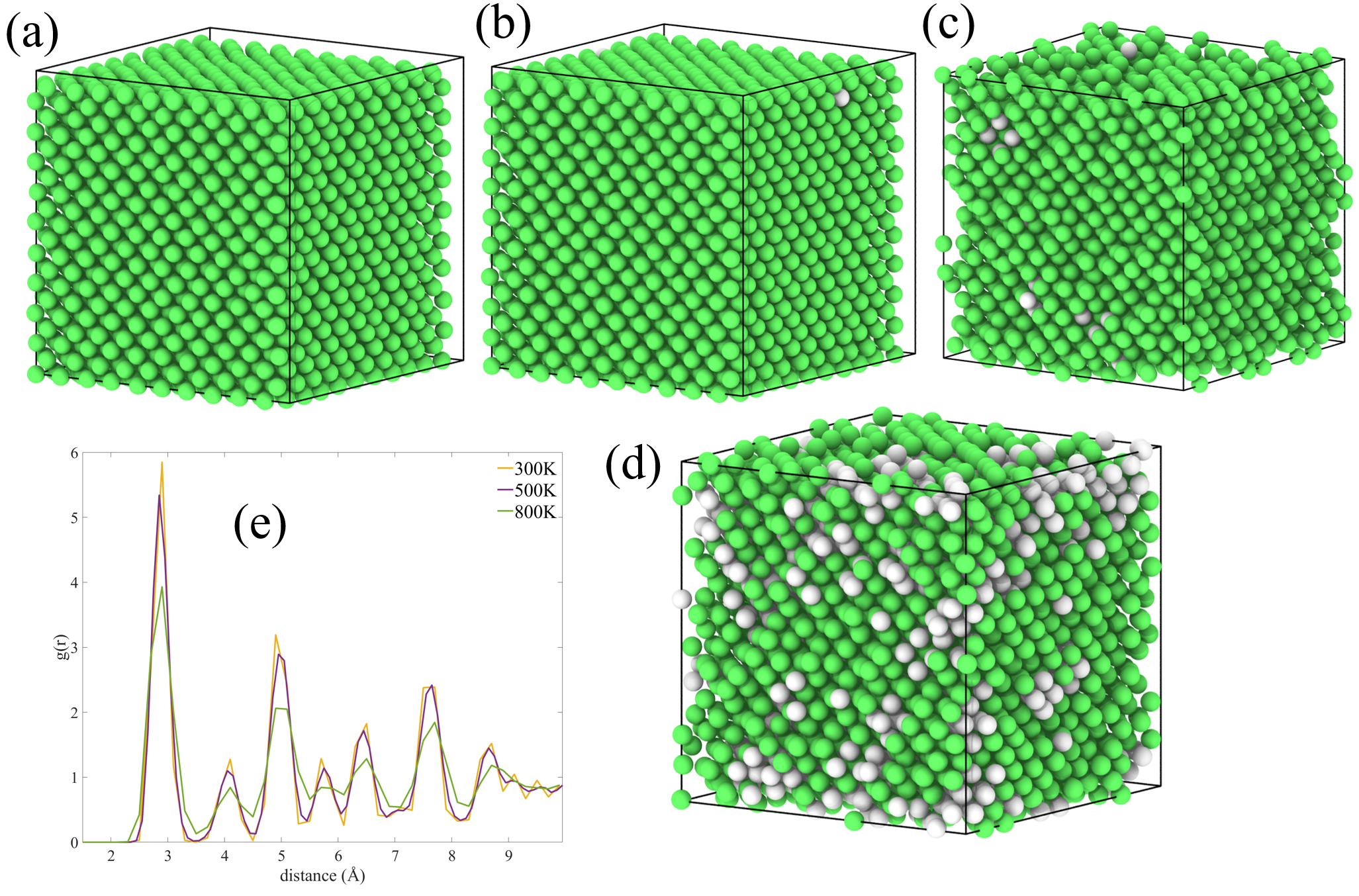}
    \caption{(a–d) Atomic configurations of bulk aluminum at increasing temperatures showing the progression from a perfect FCC lattice to thermally disordered states. (e) Corresponding radial distribution function $g(r)$ at 300~K, 500~K, and 800~K.}
    \label{fig:structure}
\end{figure}

The initial structure was a face-centered cubic (FCC) lattice of pure aluminum comprising
$10 \times 10 \times 10$ conventional unit cells (4000 atoms) under fully periodic boundary
conditions. The equilibrium lattice constant was set to $a_0 = 4.05~\text{\AA}$, consistent
with experimental data for aluminum near room temperature. All simulations were carried out in the isothermal--isobaric ensemble (NPT) using a
Nosé--Hoover thermostat and barostat \cite{hoover1985canonical} to control the system
temperature and pressure. This ensemble allows both atomic positions and the simulation
cell volume to fluctuate, enabling equilibration of the material density at each target
temperature under approximately ambient pressure conditions. The equations of motion
were integrated using the velocity--Verlet algorithm with a time step of 1~fs, which
ensures accurate resolution of atomic vibrations and stable integration of the equations
of motion. Independent equilibration simulations were conducted at temperatures of 5~K, 100~K,
300~K, 500~K, and 800~K. Each system was equilibrated for 50~ps to allow full relaxation
of both atomic coordinates and cell dimensions, followed by a 100~ps production run,
during which atomic coordinates, velocities, and thermodynamic quantities were recorded
at regular intervals of 100~fs. This temperature range captures the progressive thermal
disordering of the crystal lattice, as illustrated in
Figure~\ref{fig:structure}(a--d).

At lower temperatures, the system maintains a highly ordered crystalline configuration,
while at elevated temperatures atoms exhibit increased vibrational amplitudes and partial
loss of long-range order due to anharmonic effects. The corresponding radial distribution
functions, $g(r)$, shown in Figure~\ref{fig:structure}(e), confirm the retention of
crystalline order at low temperatures and the broadening of coordination peaks at higher
temperatures, consistent with thermal expansion and increased atomic mobility.

\subsubsection{Data Extraction and Preprocessing.}
For each temperature, the trajectory data were collected in terms of atomic positions $\mathbf{r}_i = (x_i, y_i, z_i)$ and velocities $\mathbf{v}_i = (v_{x,i}, v_{y,i}, v_{z,i})$, along with the simulation temperature $T$ and integration timestep $\Delta t$. 
Periodic boundary conditions were treated using the minimum image convention to preserve the continuity of atomic motion across cell boundaries. 
The resulting dataset consisted of multiple equilibrium configurations spanning the selected temperature range, each representing the thermally equilibrated microstate of the aluminum crystal. 
These configurations were subsequently used to construct graph-based atomic representations for training the GNN model described in Section~2.2.

\begin{table}[h!]
\centering
\caption{Summary of molecular dynamics simulation parameters used for dataset generation.}
\begin{tabular}{lll}
\hline
\textbf{Parameter} & \textbf{Description} & \textbf{Value} \\ 
\hline
Material & Pure Aluminum & --- \\
Potential & Embedded Atom Method (EAM) & --- \\
Lattice type & Face-centered cubic (FCC) & $a_0 = 4.05~\text{\AA}$ \\
Number of atoms & $4000$ & ($10\times10\times10$ unit cells) \\
Boundary conditions & Periodic in all directions & --- \\
Ensemble & NPT  & --- \\
Integration algorithm & Velocity–Verlet & --- \\
Timestep & 1~fs & --- \\
Temperature range & 5–800~K & (5 temperatures) \\
Equilibration time & 50~ps & --- \\
Production time & 100~ps & --- \\
Data sampling interval & 100~fs & --- \\
\hline
\end{tabular}
\label{tab:MDparams}
\end{table}

\noindent
The above simulation workflow ensures that all atomic trajectories are physically consistent with the underlying many-body potential and provide statistically independent configurations across temperatures. 
These equilibrated datasets form the foundation for developing the data-driven atomic trajectory prediction framework described in the subsequent sections.
\subsection{Graph Neural Network Architecture}

To capture spatial interactions among atoms, we represent each atomic configuration as a graph, with atoms as nodes and edges connecting neighboring atoms. A new graph is constructed independently at each simulation timestep, based on the atomic positions at that moment. Each graph is processed by a stacked message passing architecture along with an attention mechanisms. We organize the description of the model into five components: (1) input representation of features, (2) transformation of those features using node and edge embeddings, (3) message passing and attention layers, (4) displacement prediction, and (5) position update mechanism. Each is described in detail below.

\subsubsection{Input Representation}
\paragraph{Node Features:}
We represent each atomic configuration at a given timestep as a graph \( G = (V, E) \), where \( V \) is the set of atoms and \( E \) is the set of pairwise connections between atoms that lie within a fixed cutoff distance. A new graph is constructed at each simulation timestep based on the current atomic positions.

Each node \( i \in V \) corresponds to an atom and is associated with a set of scalar features. The input feature vector for atom \( i \) is given by:
\[
\mathbf{x}_i = [x_i, y_i, z_i, v_{x,i}, v_{y,i}, v_{z,i}, \mathrm{CSP}_i, \mathrm{CNA}_i, \mathrm{MYQ}_{1,i}, \mathrm{MYQ}_{2,i}, \mathrm{CN}_i, \Delta t, T]
\]

Here, \( (x_i, y_i, z_i) \) are the Cartesian coordinates of atom \( i \), and \( (v_{x,i}, v_{y,i}, v_{z,i}) \) are its velocity components. \( \mathrm{CSP}_i \) (centro-symmetry parameter), \( \mathrm{CNA}_i \) (common neighbor analysis), \( \mathrm{MYQ}_{1,i} \) and \( \mathrm{MYQ}_{2,i} \) (first and second components of the Minkowski structure metric), and \( \mathrm{CN}_i \) (coordination number) encode local structural environments around each atom. The scalar \( \Delta t \) is the time interval between simulation steps, and \( T \) is the system temperature. Both \( \Delta t \) and \( T \) are included as global context and are broadcast to all nodes.

In addition, each atom has a discrete atomic number \( z_i \), which identifies the element type. This is mapped to a learnable vector representation through a trainable embedding layer.

\paragraph{Edge Features:}

For each edge \( (i, j) \in E \), we compute a set of relative features that describe the difference in position and velocity between the connected atoms:
\[
\mathbf{e}_{ij} = [x_j - x_i, y_j - y_i, z_j - z_i, v_{x,j} - v_{x,i}, v_{y,j} - v_{y,i}, v_{z,j} - v_{z,i}]
\]
These features capture both the direction and magnitude of relative atomic motion and are computed using \emph{wrapped coordinates} under \emph{periodic boundary conditions (PBCs)}. Consequently, atoms that cross a simulation box boundary (e.g., in the $+x$ direction) during equilibration at a given pressure and temperature are mapped back into the simulation cell from the opposite face (i.e., the $-x$ direction).

The Euclidean distance between atoms \( i \) and \( j \) is computed as:
\[
r_{ij} = \|\mathbf{r}_j - \mathbf{r}_i\| = \sqrt{(x_j - x_i)^2 + (y_j - y_i)^2 + (z_j - z_i)^2}
\]

\paragraph{RBF Encoding}

We use a radial basis function (RBF) expansion to map scalar inputs, specifically interatomic distances, time intervals, and temperature, into smooth, multi-dimensional vectors. This allows the model to incorporate continuous scalar information in a form suitable for neural processing. For interatomic distances, the expansion is defined as:

\[
\psi_{\text{rbf}}(r_{ij}) = \exp\left[-\gamma (r_{ij} - c_k)^2\right], \quad k = 1, \ldots, N_{\text{rbf}},
\]

where \(\gamma\) is a fixed width parameter, and \(c_k\) are the RBF centers. The centers are linearly spaced in the interval \([0, R_{cutoff}]\), where the cutoff, $R_{cutoff}$, is set to \(6.287\,\text{\AA}\), matching the radius used for graph construction and neighbor selection. We use \(N_{\text{rbf}} = 32\) and set \(\gamma = 1/\Delta^2\), where \(\Delta\) is the spacing between adjacent centers.

The same RBF formulation is applied to encode the time interval \(\Delta t\) and temperature \(T\), using separate sets of 32 centers in each case. These centers are linearly spaced over appropriate input ranges, chosen to include the values of \(\Delta t\) and \(T\) seen in the training data. In all cases, we use the same definition for \(\gamma\), based on the spacing between centers.

In the remainder of the paper, we refer to this transformation abstractly as \(\psi_{\text{rbf}}(\cdot)\).

\subsubsection{Feature Construction}

\paragraph{Static Node Embedding}

Each atom is associated with a set of static features that do not change during the message passing process. These include the atomic number $z_i$, the time interval $\Delta t$, the temperature $T$, and five structural descriptors: centro-symmetry parameter (CSP), common neighbor analysis (CNA), two Minkowski structure metrics (MYQ[1], MYQ[2]), and coordination number (CN). 

The atomic number $z_i$ is mapped to a learnable embedding vector $\psi_{\text{emb}}(z_i) \in \mathbb{R}^{64}$. The scalar quantities $\Delta t$ and $T$ are expanded using radial basis function (RBF) encodings $\psi_{\text{rbf}}(\Delta t)$ and $\psi_{\text{rbf}}(T)$, both in $\mathbb{R}^{32}$. The five structural features are concatenated into a vector $\mathbf{f}_i \in \mathbb{R}^{5}$ and passed through a shared linear layer followed by a GELU activation to produce a 64-dimensional vector $\psi_{\text{struct}}(\mathbf{f}_i) \in \mathbb{R}^{64}$.

The resulting static node embedding $\mathbf{h}_i^{\text{static}} \in \mathbb{R}^{192}$ is formed by concatenating all components:
\[
\mathbf{h}_i^{\text{static}} = 
\left[
\psi_{\text{emb}}(z_i) \,\|\, 
\psi_{\text{rbf}}(\Delta t) \,\|\, 
\psi_{\text{rbf}}(T) \,\|\, 
\psi_{\text{struct}}(\mathbf{f}_i)
\right]
\]

\paragraph{Final Node and Edge Vectors}

The final input node feature vector $\mathbf{x}_i \in \mathbb{R}^{198}$ is constructed by concatenating the current atomic position $\mathbf{r}_i \in \mathbb{R}^3$, velocity $\mathbf{v}_i \in \mathbb{R}^3$, and the static node embedding $\mathbf{h}_i^{\text{static}} \in \mathbb{R}^{192}$:
\[
\mathbf{x}_i = 
\left[
\mathbf{r}_i \,\|\, 
\mathbf{v}_i \,\|\, 
\mathbf{h}_i^{\text{static}}
\right]
\]

For each edge $(i,j)$, the edge feature vector $\mathbf{e}_{ij} \in \mathbb{R}^{96}$ is constructed by concatenating two components: (1) an RBF encoding of the interatomic distance $\|\mathbf{r}_i - \mathbf{r}_j\|$, and (2) a nonlinear transformation of the relative velocity $\mathbf{v}_j - \mathbf{v}_i$:
\[
\mathbf{e}_{ij} =
\left[
\psi_{\text{rbf}}(\|\mathbf{r}_i - \mathbf{r}_j\|) \,\|\, 
\psi_{\text{vel}}(\mathbf{v}_j - \mathbf{v}_i)
\right]
\]
where $\psi_{\text{rbf}}(\cdot) \in \mathbb{R}^{32}$ is defined as in the RBF Encoding section, and $\psi_{\text{vel}}(\cdot) \in \mathbb{R}^{64}$ is a two-layer MLP with GELU activation applied to the relative velocity.

The resulting dimensions of the node and edge vectors are summarized in Table~\ref{tab:node_edge_features}.
\begin{table}[h]
\centering
\caption{Summary of node and edge features used as model input.}
\begin{tabular}{l l c}
\hline
\multicolumn{3}{c}{\textbf{Node features}} \\
\hline
\textbf{Feature} & \textbf{Transformation} & \textbf{Dimension} \\
\hline
Atomic position $(x_i, y_i, z_i)$ & Raw coordinates & 3 \\
Atomic velocity $(v_{x,i}, v_{y,i}, v_{z,i})$ & Raw components & 3 \\
Atomic number $z_i$ & Trainable embedding & 64 \\
Time interval $\Delta t$ & RBF encoding & 32 \\
Temperature $T$ & RBF encoding & 32 \\
\hline
Centro-symmetry parameter (CSP) & \multirow{5}{*}{Shared linear layer + GELU} & \multirow{5}{*}{64 (total)} \\
Common neighbor analysis (CNA) &  &  \\
Minkowski structure metric (MYQ[1]) &  &  \\
Minkowski structure metric (MYQ[2]) &  &  \\
Coordination number (CN) &  &  \\
\hline
\textbf{Total} & & \textbf{198} \\
\\[-1.2ex]
\hline
\multicolumn{3}{c}{\textbf{Edge features}} \\
\hline
\textbf{Feature} & \textbf{Transformation} & \textbf{Dimension} \\
\hline
Interatomic distance $\|\mathbf{r}_i - \mathbf{r}_j\|$ & RBF encoding & 32 \\
Relative velocity $\mathbf{v}_j - \mathbf{v}_i$ & MLP (2-layer with GELU) & 64 \\
\hline
\textbf{Total} & & \textbf{96} \\
\hline
\end{tabular}
\label{tab:node_edge_features}
\end{table}

\subsubsection{Graph Architecture Overview}

\paragraph{Block Structure and Stacking}

The model is composed of a sequence of identical blocks, each consisting of message passing and self-attention layers. Within each block, node features are updated through two stages: first via message passing over neighboring nodes, and then through an attention-based refinement mechanism inspired by transformer architectures. Each block also includes a residual connection, layer normalization, and a feedforward network (FFN) to stabilize training and improve expressiveness.

After computing refined node features $\mathbf{h}_i^{\text{attn}}$ within a block, the model predicts a displacement vector $\Delta \mathbf{r}_i$ for each atom. The position is updated immediately:
\[
\mathbf{r}_i^{t + \Delta t} = \mathbf{r}_i^{t} + \Delta \mathbf{r}_i
\]
This updated position is then used to reconstruct the graph for the next block. In this way, the model performs position updates \emph{internally} after each block rather than predicting only at the final layer. This iterative refinement enables the network to learn and correct atomic displacements in stages. A total of $N = 6$ blocks are stacked.

\paragraph{Graph Construction per Timestep}

At each simulation timestep, a new graph is constructed based solely on the current atomic positions. Atoms are treated as nodes, and edges are defined between atoms that lie within a fixed cutoff distance. This graph captures the spatial configuration of the system at that specific moment.

Importantly, graph construction is performed \textit{independently} at each timestep, without carrying over information from previous frames. As a result, the model is not a temporal sequence model; it performs per-timestep inference. To compensate for this, the time interval $\Delta t$ is provided as a global node feature, allowing the model to learn how far into the future it is expected to predict.

Each constructed graph is static during forward propagation through the GNN layers within a block, but is updated between blocks based on the newly predicted atomic positions.

\subsubsection{Message Passing and Attention Mechanism}

\paragraph{Message Passing : }

The message passing layers follow a standard formulation where each node exchanges information with its neighbors to update its state. For a given node $i$, messages are computed from its neighboring nodes $j \in \mathcal{N}(i)$ and the associated edge features $\mathbf{e}_{ij}$. The message from node $j$ to node $i$ is given by:
\[
\mathbf{m}_{ij} = \text{MLP}_m([\mathbf{h}_i, \mathbf{h}_j, \mathbf{e}_{ij}])
\]
where $\text{MLP}_m$ is a two-layer perceptron with GeLU activation, and $[\cdot]$ denotes vector concatenation. The incoming messages are averaged:
\[
\mathbf{a}_i = \frac{1}{|\mathcal{N}(i)|} \sum_{j \in \mathcal{N}(i)} \mathbf{m}_{ij}
\]
The updated node feature is then computed as:
\[
\mathbf{h}_i' = \text{MLP}_u([\mathbf{h}_i, \mathbf{a}_i])
\]
where $\text{MLP}_u$ is another two-layer network with GeLU activation. The output $\mathbf{h}_i'$ serves as the input to the attention mechanism in the next stage.

\paragraph{Self-Attention Mechanism :}

After message passing, node representations are refined using a self-attention mechanism adapted to graph-structured data. For each node $i$, a query vector is computed from the current node state, while keys and values are computed from neighboring node features and edge features:
\[
\mathbf{q}_i = W_q \mathbf{h}_i, \quad 
\mathbf{k}_{ij} = W_k^h \mathbf{h}_j + W_k^e \mathbf{e}_{ij}, \quad 
\mathbf{v}_{ij} = W_v^h \mathbf{h}_j + W_v^e \mathbf{e}_{ij}
\]
The attention weight is computed using a scaled dot-product:
\[
\alpha_{ij} = \frac{\exp(\mathbf{q}_i^\top \mathbf{k}_{ij} / \sqrt{d})}{\sum_{k \in \mathcal{N}(i)} \exp(\mathbf{q}_i^\top \mathbf{k}_{ik} / \sqrt{d})}
\]
The attention-weighted sum of values produces the message vector:
\[
\mathbf{m}_i = \sum_{j \in \mathcal{N}(i)} \alpha_{ij} \mathbf{v}_{ij}
\]

\paragraph{Residual Connections and Normalization : }

The attention output $\mathbf{m}_i$ is added to a linear projection of the input $\mathbf{x}_i$, followed by layer normalization:
\[
\mathbf{h}_i' = \text{LayerNorm}\left( \text{Linear}(\mathbf{x}_i) + \mathbf{m}_i \right)
\]
Then, a feedforward network (FFN) is applied with a second residual connection and normalization:
\[
\mathbf{h}_i^{\text{attn}} = \text{LayerNorm}\left( \mathbf{h}_i' + \text{FFN}(\mathbf{h}_i') \right)
\]
The FFN consists of two linear layers with GeLU activation in between. These residual paths and normalization layers help stabilize training and allow for deeper stacking of attention blocks.

\subsubsection{Prediction Layer}

\paragraph{Position Update and Iteration}

The model applies a sequence of attention blocks. At the input of each block, the current atomic positions $\mathbf{r}_i$ are concatenated with atomic velocities and static node features to form the node feature vector. After message passing and attention within a block, the resulting node features $\mathbf{h}_i^{\text{attn}}$ are projected to a displacement vector using a linear layer:
\[
\Delta \mathbf{r}_i = W_{\text{out}} \mathbf{h}_i^{\text{attn}}
\]
This displacement is immediately added to the current atomic positions:
\[
\mathbf{r}_i \leftarrow \mathbf{r}_i + \Delta \mathbf{r}_i
\]
The updated positions are then used to construct the node feature input for the next attention block. This procedure is repeated for all blocks in the stack. iterative application of attention blocks results in a sequence of incremental position updates within a single forward pass.

\paragraph{Final Output}

After all $N$ attention blocks have been applied, the model outputs the final predicted atomic positions. At each block, a displacement vector $\Delta \mathbf{r}_i^{(b)}$ is computed and immediately added to the current atomic position. This results in a cumulative update applied sequentially over the $N$ blocks:
\[
\mathbf{r}_i^{t + \Delta t} = \mathbf{r}_i^t + \sum_{b=1}^{N} \Delta \mathbf{r}_i^{(b)}
\]
The final position $\mathbf{r}_i^{t + \Delta t}$, after all updates, serves as the model’s output.

\subsection{Training Setup}
We constructed our graphs for training and validation as part of the preprocessing step. Atomic states from raw dump files were extracted using a lazy-loading indexer to handle large-scale trajectory data efficiently. To prevent data leakage, we employed a temporal partitioning strategy. For each simulation trajectory, the first $80\%$ of time steps are allocated to the training set, while the final $20\%$ are reserved for validation.

We implemented our model using PyTorch and PyTorch Geometric modules. Training was done in a distributed computing setup using \emph{DistributedDataParallel} to leverage multi-GPU capabilities. The data loading pipeline utilized DistributedSampler to enforce data partitions. We used a per-GPU batch size of 4. We employed 4 worker processes per GPU for increased throughput. 

The network parameters were optimized using the AdamW optimizer with an initial learning rate of $\eta = 1 \times 10^{-3}$ and a weight decay of $1 \times 10^{-5}$ for regularization. We utilized a learning rate scheduler (\emph{ReduceLROnPlateau}) that monitors the validation loss, decaying the learning rate by a factor of 0.5 if the loss fails to improve for 5 consecutive epochs. To stabilize training, we applied gradient clipping with a maximum norm of 1.0. 

The loss function was defined as the Mean Squared Error (MSE) between predicted and true positions:
\[
\mathcal{L} = \frac{1}{N} \sum_i \|\mathbf{r}_i^{(\text{pred})} - \mathbf{r}_i^{(\text{true})}\|^2
\]
To facilitate reproducibility, we fixed the random seeds for PyTorch, NumPy, and the Python random number generator. We saved checkpoints at every epoch of the training run. The best performing model was selected based on the lowest average loss on the validation set.


\begin{table}[h]
\centering
\caption{Hyperparameters and structural details of the model used in this work.}
\begin{tabular}{lll}
\hline
\textbf{Parameter} & \textbf{Value} & \textbf{Description} \\
\hline
$D_{\text{input}}^{\text{node}}$ & 13 & Raw input features per node\\
$D_{\text{hidden}}$ & 128 & Main hidden dimension (Attention/FFN input) \\
$D_{\text{input}}^{\text{edge}}$ & 6 & Raw edge features (Relative $R_{ij}$ and $v_{ij}$) \\
$D_{\text{embed}}^{\text{edge}}$ & 96 & Edge embedding dimension ($32_{\text{RBF}} + 64_{\text{MLP}}$) \\
$N_{\text{rbf}}$ & 32 & Number of RBF centers for Dist, Time, and Temp \\
$R_{\text{cutoff}}$ & 6.287 A& Spatial cutoff radius for graph edges \\
$T_{\text{cutoff}}$ & 10.0 fs& Temporal lookback cutoff for node history \\
$N_{\text{blocks}}$ & 6 & Number of Message Passing Attention blocks \\
$N_{\text{heads}}$ & 8 & Number of attention heads \\
$D_{\text{head}}$ & 16 & Dimension per attention head ($D_{hidden}/N_{heads})$\\
$N_{\text{species}}$ & 1& Size of atomic species embedding table \\
FFN Width& 512 & Hidden units in block feedforward layers ($4 \times D_{\text{hidden}}$) \\
Output dimension & 3 & Predicted position update ($\Delta x, \Delta y, \Delta z$) \\
Temp. embedding & RBF (32) & Embedding of system Temperature\\
\hline
\end{tabular}
\label{tab:alloymd_model_params}
\end{table}


\section{Results and Discussion}

\subsection{Atomic Coordinate Prediction and Trajectory Fidelity}

\begin{figure}[H]
\centering
\includegraphics[width=1\textwidth]{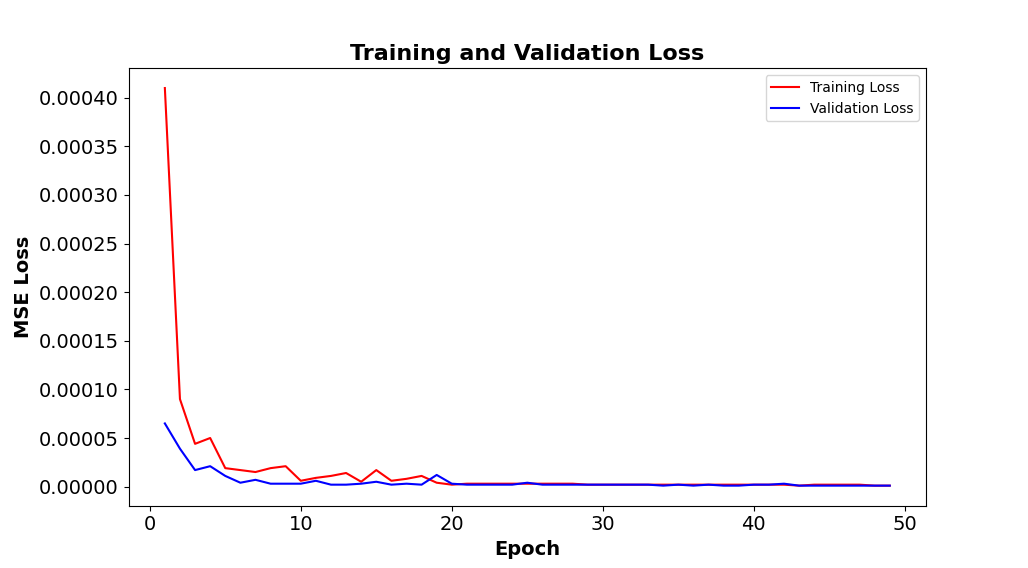}
\caption{Training and validation loss as a function of epoch.}
\label{fig:loss}
\end{figure}

The training process exhibited rapid convergence and high numerical stability, indicating that the proposed graph neural network effectively learns the underlying atomic evolution operator for the aluminum system. As shown in Fig.~\ref{fig:loss}, the model minimizes the loss function within the first ten epochs and subsequently reaches a stable plateau. The validation loss decreases from 0.000065 to 0.000001 over the 50-epoch training run, corresponding to an average displacement error of approximately 0.001~\AA. The close overlap between the training and validation loss curves indicates strong generalization within the training distribution and an absence of overfitting. The lack of divergence in validation loss further suggests that the model architecture is well matched to the complexity of the molecular dynamics trajectory data. Beyond interpolation, an important test of the model’s utility is its ability to predict atomic positions at time intervals extending beyond the temporal horizon used during training. The model was trained using trajectories with $\Delta t \leq 10$, and inference was performed up to $\Delta t = 50$ at 350~K to assess extrapolation behavior. As summarized in Table~\ref{tab:delta_t}, the model achieves near-perfect reconstruction within the interpolation regime, with displacement errors remaining below 0.002~\AA. When extrapolating beyond the training horizon, a distinct increase in error is observed at $\Delta t = 15$, where the displacement error increases to 0.033~\AA. This behavior is expected, as the model is operating outside its training distribution. Importantly, the model does not exhibit catastrophic divergence. At $\Delta t = 50$, the mean displacement error is 0.13~\AA, which corresponds to a relative error of approximately 4.5\% when compared to the nearest-neighbor distance in aluminum of $\sim$2.86~\AA. The displacement error increases approximately linearly with $\Delta t$ over this range, with an $R^2$ value of 0.97, indicating stable long-horizon behavior. Assuming this trend persists, the model error would remain below 10\% of the nearest-neighbor distance up to $\Delta t \approx 95$, suggesting that the learned surrogate captures the dominant atomic dynamics well beyond the interpolation regime.

\begin{table}[H]
    \centering
       \caption{Model performance across different $\Delta t$ at 350~K.}
    \begin{tabular}{ccc}
       \hline
         TimeStep & MSE Loss & in Angstroms \\
       \hline
         1  & 0.000000046677 & 0.00022 \\
         5  & 0.00000042483  & 0.00065 \\
         10 & 0.00000357     & 0.0019 \\
         15 & 0.0011         & 0.033 \\
         20 & 0.0024         & 0.049 \\
         30 & 0.0063         & 0.079 \\
         40 & 0.0114         & 0.118 \\
         50 & 0.0169         & 0.13 \\
       \hline
    \end{tabular}
 
    \label{tab:delta_t}
\end{table}

\begin{figure}[H]
    \centering
    \includegraphics[width=0.75\linewidth]{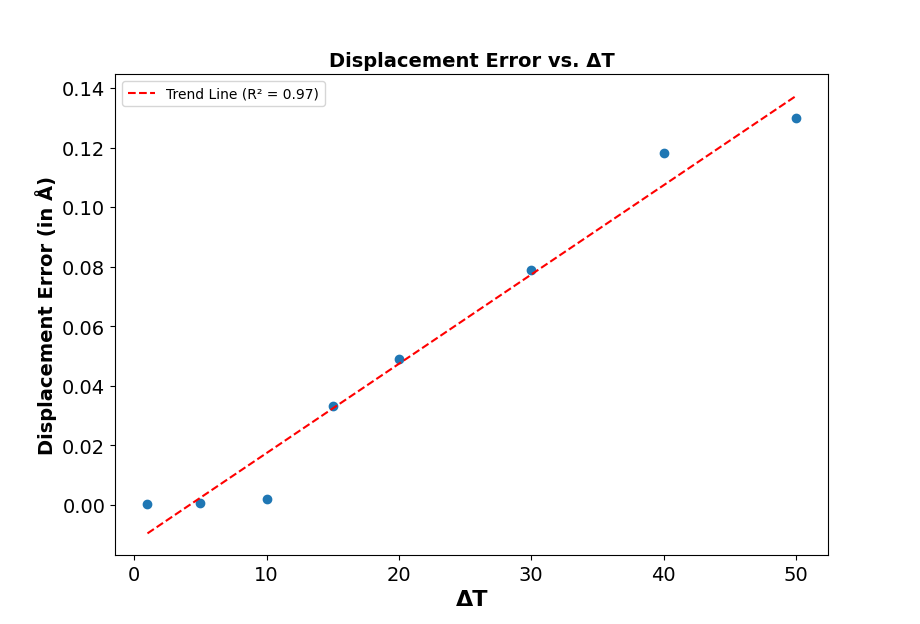}
    \caption{Model displacement error (in \AA) as a function of $\Delta t$ at 350~K.}
    \label{fig:displacement_trend}
\end{figure}

\subsection{Thermodynamic Consistency: Temperature and Volume}

In addition to reproducing atomic coordinates and local structural metrics, a physically
meaningful surrogate model must preserve global thermodynamic observables. To assess
whether the proposed GNN surrogate maintains thermodynamic consistency, we compare
system-level temperature and volume obtained from MD simulations
and GNN-predicted configurations. For this analysis, atomic configurations were sampled after $50{,}000$ MD timesteps, corresponding to a well-equilibrated state under NPT conditions. At this fixed timestep, the GNN-predicted atomic positions were used to compute the instantaneous system temperature and simulation cell volume using the same post-processing procedures applied to the reference MD trajectories. This ensures a direct, state-to-state comparison between
MD and model-derived quantities.

Table~\ref{tab:temp_volume} summarizes the comparison at representative low and high
temperatures (300~K and 800~K). The GNN surrogate reproduces both temperature and
volume with high fidelity across this range. At 300~K, the predicted temperature differs
from the MD reference by less than 1~K, while the volume mismatch is below $0.01\%$.
Similarly, at 800~K, the temperature deviation remains under 2~K, and the predicted
thermal expansion is accurately captured.

\begin{table}[h]
\centering
\caption{Comparison of actual MD-derived and GNN-predicted system temperature and
volume at $50{,}000$ timesteps.}
\label{tab:temp_volume}
\begin{tabular}{cccc}
\hline
Nominal Temperature & Quantity & MD & GNN \\
\hline
\multirow{2}{*}{300 K}
& Temperature (K) & 298.6 & 299.3 \\
& Volume (\AA$^{3}$) & 60373 & 60375 \\
\hline
\multirow{2}{*}{800 K}
& Temperature (K) & 793.5 & 791.7 \\
& Volume (\AA$^{3}$) & 63960 & 63984 \\
\hline
\end{tabular}
\end{table}

The close agreement in both temperature and volume demonstrates that the learned
surrogate does not introduce spurious heating, cooling, or volumetric drift when propagating
atomic configurations. This result is particularly significant because the model directly
predicts atomic displacements without explicit force evaluation or thermostat coupling.
Despite this, the GNN preserves the thermodynamic state encoded in the MD data,
indicating that it has learned a physically consistent representation of the underlying
many-body dynamics. Together with the coordinate-level accuracy reported in Section~3.1 and the structural
and dynamical validations presented in Sections~3.3 and 3.4, these results confirm that
the GNN surrogate functions as a stable and thermodynamically faithful alternative to
classical MD integration within the validated regime.

\subsection{Structural Fidelity via Radial Distribution Functions}

\begin{figure}[H]
\centering
\includegraphics[width=1\linewidth]{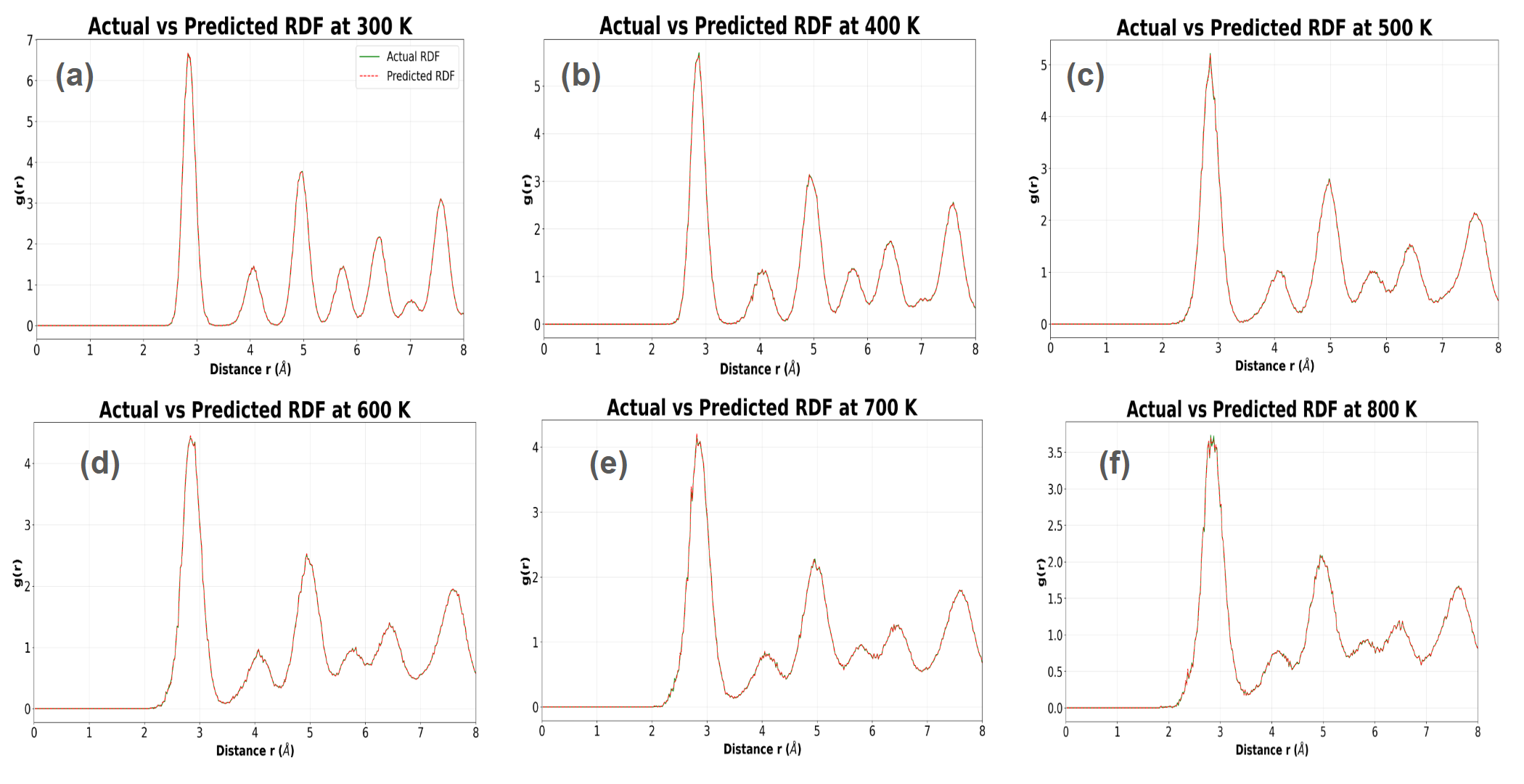}
\caption{Actual versus model-derived radial distribution functions at 300--800~K for $\Delta t = 5$.}
\label{fig:rdf_dt5}
\end{figure}

\begin{figure}[H]
\centering
\includegraphics[width=1\linewidth]{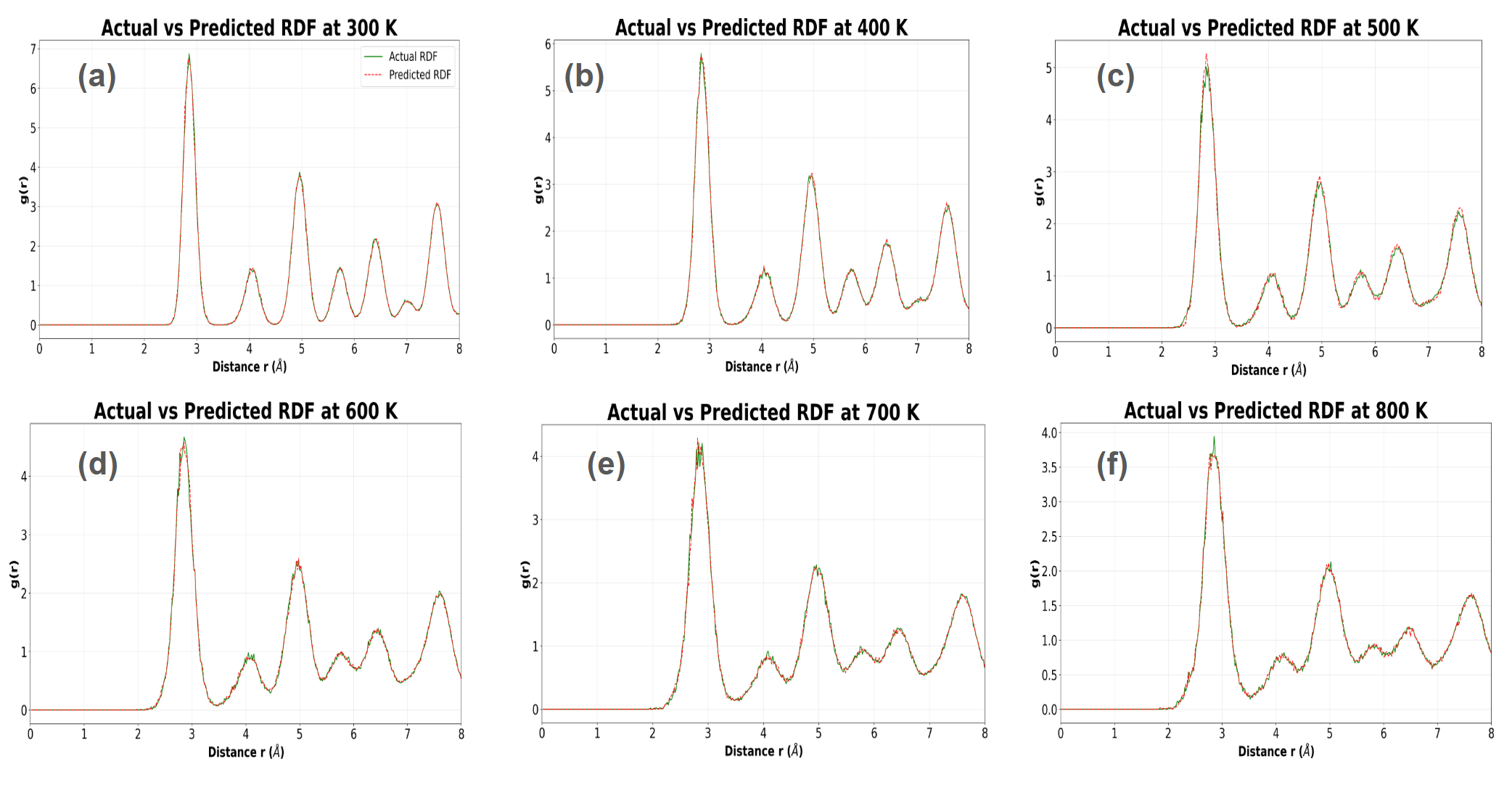}
\caption{Actual versus model-derived radial distribution functions at 300--800~K for $\Delta t = 50$.}
\label{fig:rdf_dt50}
\end{figure}

To assess whether the predicted atomic coordinates preserve physically meaningful local structure, we compared the radial distribution functions (RDFs), $g(r)$, computed from the model-predicted configurations against those obtained directly from molecular dynamics simulations. Figure~\ref{fig:rdf_dt5} shows the RDF comparison across temperatures ranging from 300~K to 800~K for $\Delta t = 5$. Across this temperature range, the predicted RDFs show strong overlap with the ground-truth MD data, accurately capturing both the peak locations and the progressive peak broadening associated with increased thermal disorder. This agreement demonstrates that the model does not merely reproduce atomic positions in a pointwise sense but also preserves the underlying pairwise spatial correlations that define the crystal structure. Figure~\ref{fig:rdf_dt50} presents the RDFs at $\Delta t = 50$, corresponding to the long-horizon extrapolation regime. Despite the accumulation of absolute position error at larger $\Delta t$, the RDF structure remains well preserved. The first, second, and third coordination peaks are accurately reproduced in both position and intensity. This indicates that although atoms experience modest drift over long time intervals, the model maintains short-range order and avoids unphysical artifacts such as atomic overlap or lattice distortion. Notably, the predicted probability density correctly decays to zero beyond the first few coordination shells, indicating that the model has learned both physically allowed and forbidden atomic configurations.

\subsection{Dynamical Consistency via Mean Squared Displacement}

\begin{figure}[H]
\centering
\includegraphics[width=1\linewidth]{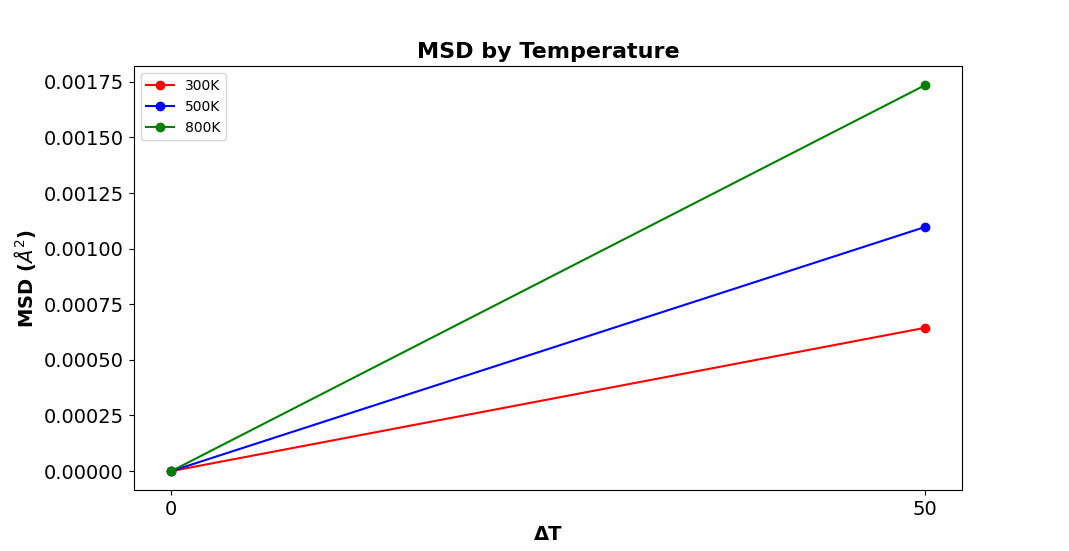}
\caption{Mean squared displacement as a function of time steps at 300~K, 500~K, and 800~K for $\Delta t = 50$.}
\label{fig:msd}
\end{figure}

To further evaluate the dynamical fidelity of the learned surrogate, we analyzed the mean squared displacement (MSD) computed from model-predicted trajectories at 300~K, 500~K, and 800~K for $\Delta t = 50$. As shown in Fig.~\ref{fig:msd}, the MSD increases monotonically with time for all temperatures, with the highest MSD observed at 800~K and the lowest at 300~K. This temperature-dependent ordering is consistent with physical expectations and reflects the increased vibrational amplitude and atomic mobility at elevated temperatures. The model accurately captures both the relative magnitude and the qualitative trend of the MSD across temperatures, indicating that it has learned the relationship between thermal energy and atomic motion rather than simply memorizing equilibrium configurations. The preservation of correct MSD behavior, together with the RDF results, demonstrates that the GNN surrogate reproduces both structural and dynamical signatures of the underlying molecular dynamics simulations, even when operating beyond the temporal horizon used during training.

\begin{table}[H]
    \centering
      \caption{Model performance across temperature range.}
    \begin{tabular}{ccc}
    \hline
         Temp (in K) & MSE & in Angstroms \\
    \hline
         5   & 0.000000218 & 0.0004669047012 \\
         50  & 0.000000843 & 0.0009181503145 \\
         100 & 0.000000306 & 0.0005531726674 \\
         150 & 0.000000466 & 0.0006826419266 \\
         200 & 0.000000186 & 0.0004312771731 \\
         300 & 0.000000431 & 0.0006565059025 \\
         350 & 0.000000425 & 0.0006519202405 \\
         400 & 0.000000577 & 0.0007596051606 \\
         450 & 0.000000282 & 0.0005310367219 \\
         500 & 0.000000519 & 0.0007204165462 \\
         550 & 0.000000438 & 0.0006618156843 \\
         600 & 0.000000399 & 0.0006316644679 \\
         650 & 0.000000683 & 0.0008264381405 \\
         700 & 0.000000666 & 0.0008160882305 \\
         750 & 0.000000709 & 0.0008420213774 \\
         800 & 0.000000703 & 0.0008384509527 \\
    \hline
    \end{tabular}
  
    \label{tab:temp}
\end{table}

\section{Ablation Study}

In the context of machine learning, an ablation study refers to the systematic removal or modification of components of a model to assess their contribution to predictive performance. This usage is distinct from physical ablation processes in materials science, such as erosion or mass loss under extreme thermal or mechanical loading. In this study, ablation is employed purely as an analysis tool to understand how architectural choices in the graph neural network influence its ability to reproduce atomistic trajectories. Specifically, we examine the effects of network depth, defined by the number of message-passing and attention blocks, and network width, defined by the hidden feature dimension, on the accuracy of predicted atomic displacements.

\subsection{Effect of Network Depth}

We first examine the sensitivity of model performance to network depth by varying the number of stacked message-passing and attention blocks from six to two while keeping all other hyperparameters fixed. The results are summarized in Table~\ref{tab:placeholder}. Reducing the depth from six to four blocks leads to only a marginal increase in the displacement error, from $0.0008$~\AA\ to $0.0009$~\AA. Even with a highly reduced architecture consisting of only two blocks, the model achieves a displacement error of $0.0010$~\AA, which remains well below typical thermal vibration amplitudes in crystalline aluminum. This behavior indicates that the atomic dynamics of the FCC aluminum lattice are relatively smooth and locally correlated, allowing the dominant interactions to be captured with a limited number of message-passing steps. From a physical perspective, this suggests that the local atomic environment encoded within a few coordination shells provides sufficient information for accurate short-time trajectory prediction. From a computational standpoint, the weak dependence on depth implies that shallower architectures may be sufficient for large-scale simulations, enabling faster inference without a significant loss in accuracy.

\begin{table}[H]
    \centering
    \caption{Changes in loss when decreasing the number of message-passing blocks.}
    \begin{tabular}{ccc}
       \hline
         Num Blocks & MSE Loss & Loss in Angstroms \\
       \hline
         6 & 0.0000006 & 0.0008 \\
         4 & 0.0000009 & 0.0009 \\
         2 & 0.000001  & 0.0010 \\
       \hline
    \end{tabular}
    \label{tab:placeholder}
\end{table}

\subsection{Effect of Hidden Feature Dimension}

We next investigate the role of network width by varying the hidden feature dimension while fixing the network depth at four blocks. The results, shown in Table~\ref{tab:placeholder}, reveal a clear threshold behavior. Reducing the hidden dimension from 128 to 64 results in only a minor increase in the displacement error, from $0.0009$~\AA\ to $0.0010$~\AA, indicating that the model retains sufficient representational capacity at this reduced width. However, a further reduction to a hidden dimension of 32 leads to a substantial degradation in performance, with the displacement error increasing to $0.0030$~\AA. This sharp increase suggests that a minimum feature space dimensionality is required to adequately represent the high-dimensional phase space associated with atomic vibrations, local structural descriptors, and velocity information. In physical terms, the hidden feature dimension controls the model’s capacity to encode subtle variations in bonding geometry and thermal motion. Insufficient width limits this capacity and leads to loss of predictive fidelity. These results indicate that, while extreme depth is not necessary, maintaining an adequate feature dimension is critical for accurately capturing atomistic dynamics in metallic systems.

\begin{table}[H]
    \centering
    \caption{Change in loss when decreasing the hidden feature dimension for a four-block network.}
    \begin{tabular}{ccc}
       \hline
         Hidden Dim (4 Blocks) & MSE Loss & Loss in Angstroms \\
       \hline
         128 & 0.0000009 & 0.0009 \\
         64  & 0.000001  & 0.0010 \\
         32  & 0.000009  & 0.0030 \\
       \hline
    \end{tabular}
    \label{tab:placeholder}
\end{table}

\section{Conclusion}

In this work, we developed a graph neural network–based surrogate model that directly
predicts atomic displacements in molecular dynamics simulations of bulk aluminum,
thereby learning the atomic evolution operator and bypassing explicit force evaluation
and numerical time integration. The model was trained on classical EAM molecular
dynamics trajectories spanning a wide temperature range from 5~K to 800~K and
incorporates atomic positions, velocities, thermodynamic state variables, and local
structural descriptors within a physically motivated graph representation. Within the training horizon, the surrogate achieves sub-\AA ngstrom accuracy and
exhibits stable, approximately linear error growth during temporal extrapolation up
to $\Delta t = 50$. Beyond coordinate-level accuracy, the model preserves key physical
observables, including radial distribution functions and temperature-dependent mean
squared displacement trends, demonstrating that both local structural order and
dynamical behavior of the FCC aluminum lattice are retained even during long-horizon
prediction. Importantly, direct comparison of model-derived and molecular-dynamics reference
thermodynamic observables confirms that the GNN surrogate preserves global physical
consistency. System temperature and volume evaluated at an equilibrated state
(50{,}000 timesteps) show near-perfect agreement with MD results across the examined
temperature range, indicating the absence of spurious thermal or volumetric drift
despite the elimination of explicit force evaluation and thermostat coupling. This
thermodynamic fidelity underscores that the learned surrogate captures not only
microscopic atomic motion but also the macroscopic state encoded in the MD data. The effectiveness of the proposed approach arises from the locality of metallic bonding
and the ability of message-passing and attention mechanisms to efficiently capture
coordination-shell–level interactions. Ablation studies further reveal that accurate
short-time dynamics can be recovered using relatively shallow architectures, while a
minimum hidden feature dimension is required to represent the high-dimensional phase
space associated with atomic vibrations and thermal disorder. Together, these results
demonstrate that learned surrogate integrators can provide a computationally efficient
and physically faithful alternative to conventional molecular dynamics within a
validated regime. In contrast to machine-learned interatomic potentials, which aim to reproduce energies
and forces while retaining classical integrators, the present framework targets direct
trajectory propagation and offers a complementary pathway for accelerating atomistic
simulations when force-level fidelity is not strictly required.

Future work will extend this framework to more complex metallic systems, including
body-centered cubic and hexagonal close-packed metals as well as multi-component
alloys, where chemical disorder and varying coordination environments pose additional
challenges. Direct comparisons with force-based machine learning potentials and other
learned-dynamics approaches will be conducted to quantify accuracy, stability, and
computational efficiency across representative benchmarks. Incorporating additional
physical constraints and uncertainty quantification will be essential for improving
robustness and extending the applicability of surrogate integrators to longer
timescales and more heterogeneous materials systems.

\section*{Code Availability}

Workflows developed in this study are openly available at the following repository:
\url{https://github.com/mahata-lab/GNN-Molecular-Dynamics}.


\section*{Acknowledgments}

This work was supported by the Department of Mechanical and Electrical Engineering at Merrimack College. The author acknowledges the use of computational resources at the Massachusetts Green High Performance Computing Center (MGHPCC). This research also benefited from high performance computing allocations provided by the National Science Foundation through ACCESS (awards MAT250103 and MAT240094). Additional computational resources were supported by Argonne National Laboratory under the Director's Discretionary allocation for the project \textit{AIAlloyLW}. Further support was provided through a National Science Foundation MRI Award to Wilkes University (Award No.\ 1920129), which contributed essential computational infrastructure for this study.

\bibliographystyle{unsrt}  
\bibliography{references}

\end{document}